\documentclass[usenatbib]{mn2e}

\pdfoutput=1

\usepackage{url}
\usepackage{fixltx2e}
\usepackage{mathptmx}
\usepackage[pdftex]{graphicx}
\usepackage{color}
\usepackage{url}

\newcommand{\km}{\rm\thinspace km}

\newcommand{\s}{\rm\thinspace s}

\newcommand{\keV}{\rm\thinspace keV}

\newcommand{\kmps}{\hbox{$\km\s^{-1}\,$}}

\newcommand{\Zsun}{\hbox{$\thinspace \mathrm{Z}_{\odot}$}}

\begin{document}

\title[Velocities in the intracluster medium] {Velocity width
  measurements of the coolest X-ray emitting material in the cores of
  clusters, groups and elliptical galaxies}

\author [J.~S. Sanders \& A.~C. Fabian] {J.~S. Sanders$^{1,2}$ and
  A.~C. Fabian$^{1}$ \\
  $^1$ Institute of Astronomy, Madingley Road, Cambridge. CB3 0HA\\
  $^2$ Max-Planck-Institute f\"ur extraterrestrische Physik, 85748
  Garching, Germany.}

\maketitle

\begin{abstract}
  We examine the velocity width of cool X-ray emitting material using
  \emph{XMM-Newton} Reflection Grating Spectrometer (RGS) spectra of a
  sample of clusters and group of galaxies and elliptical galaxies.
  Improving on our previous analyses, we apply a spectral model which
  accounts for broadening due to the spatial extent of the
  source. With both conventional and Markov Chain Monte Carlo
  approaches we obtain limits, or in a few cases measurements, of the
  velocity broadening of the coolest X-ray material. In our sample, we
  include new observations targeting objects with compact, bright,
  line-rich cores. One of these, MACS\,J2229.7-2755, gives a velocity
  limit of $280\kmps$ at the 90 per cent confidence level. Other
  systems with limits close to $300\kmps$ include A\,1835, NGC\,4261
  and NGC\,4472. For more than a third of the targets we find limits
  better than $500\kmps$. HCG\,62, NGC\,1399 and A\,3112 show evidence
  for $\sim 400 \kmps$ velocity broadening.  For a smaller sample of
  objects, we use continuum-subtracted emission line surface
  brightness profiles to account for the spatial broadening. Although
  there are significant systematic errors associated with the
  technique ($\sim 150 \kmps$), we find broadening at the level of
  $280$ to $500\kmps$ in A\,3112, NGC\,1399 and NGC\,4636.
\end{abstract}


\section{Introduction}
Measuring the velocity distribution of the intracluster medium (ICM)
in clusters and groups of galaxies and in elliptical galaxies is
important for understanding how such structures form and the strength
of active galactic nucleus (AGN) feedback. Simulations of AGN feedback
imply the creation of ICM velocity widths of several hundred $\kmps$
\citep{Bruggen05,Heinz10}. In addition, sloshing of gas within the
potential well of a cluster can give rise to similar velocities
\citep{Ascasibar06}. Cosmologically generated turbulence is expected
to be a relatively small 5 to 15 per cent fraction of the thermal
energy density in the cores of galaxy clusters
\citep{Lau09,Vazza09,Vazza11},

The most direct method for measuring or placing limits on velocities
on clusters is examining X-ray emission line widths. This was first
done by examining a deep X-ray grating spectrum of A 1835 taken using
the RGS (Reflection Grating Spectrometer) instruments on
\emph{XMM-Newton} \citep{Sanders10_A1835}. We followed this up with a
study a sample of bright emission line objects
\citep{Sanders10_Broaden}.  As the RGS spectrometers are slitless
spectrometers, the extent of the source contributes to the line
width. For our most conservative analysis we assumed that the objects
were point sources, and therefore obtained upper limits on the
velocity broadening. For A\,1835 we measured a 90 per cent confidence
upper limit of $274\kmps$ and less than $500\kmps$ for several objects
in our sample. In A\,1835 the upper limit corresponds to the fraction
of energy in turbulence of less than 13 per cent of the thermal energy
density.

Other objects have recently been examined. \cite{Bulbul12} analysed
\emph{XMM} observations of A\,3112. Their most conservative upper
limit was around $900\kmps$. \cite{dePlaa12} recently examined two
elliptical galaxies. Using resonance scattering to provide lower
limits and line widths to find upper limits they obtain
$320<V_\mathrm{turb}<720\kmps$ for NGC 5044 and $140 <
V_\mathrm{turb}<540\kmps$ for NGC 5813.

In \cite{Sanders10_A1835} we went on to account for spectral
broadening due to the spatial extent of A\,1835. Using an
\emph{Chandra} image of the cluster in the Fe-L band we reduced our
limit to $\sim 180 \kmps$. In addition, \cite{Bulbul12} used
\emph{XMM} images to account for the surface brightness of A\,3112,
finding improved limits of $\sim 200\kmps$.

When analysing the sample of objects in \cite{Sanders10_Broaden}, we
also applied an experimental technique of modelling RGS spectra using
spatially-resolved \emph{Chandra} maps of the sources, comparing the
real line widths with predicted ones. This method had a number of
systematic issues: using projected quantities, not including the
uncertainties in the maps, assuming Solar abundance ratios and the
finite sizes of the bins used. It has also been pointed out by
M.~Markevitch that we did not subtract our modelled line widths in
quadrature from the observed line widths.

In this paper, we improve on our previous analyses to take account of
the spatial broadening in a robust manner. Firstly, we fit the RGS
spectra, modelling the spatial distribution of the source. For some
selected sources, we go on to use emission line profiles along the
dispersion direction, subtracting continuum emission, to include the
effects of the spatial broadening. We find a number of strong upper
limits, but for a few sources we are able to find moderate detections
of line broadening due to gas motions.

The Solar abundance ratios of \cite{AndersGrevesse89} are used in this
paper. Velocity values quoted are the Gaussian width, $\sigma$.

\section{Spectral analysis}
The line width measured in an RGS spectrum includes the intrinsic line
width from random or turbulent velocities and thermal motions.  The
thermal motion is included within the spectral model and is therefore
accounted for during spectral fitting. The non-intrinsic contribution
includes the instrumental line width (accounted for within the
response) and the spectral broadening due to the extent of the
source. The RGS instruments do not have slits, so the size of the
source broadens the spectrum by
\begin{equation}
  \Delta \lambda \approx \frac{0.139}{m} \Delta \theta \: \textrm{\AA},
\label{eqn:broad}
\end{equation}
where $m$ is the spectral order and $\Delta\theta$ is the half energy
width of the source in arcmin (note that the incorrect factor $0.124$
is given in some works, including \citealt{Brinkman98}).  In our
previous analyses we placed an upper limit on the line width, putting
a constraint on the velocity broadening.

The contribution to broadening from the spatial component is a fixed
broadening in wavelength. Velocity broadening gives a fractional
broadening in energy or frequency. In principle, if we include a
spectral model which characterises the fixed $\delta\lambda$
broadening from the extent, we can also determine the velocity
broadening independently. If the data quality is too low we can
constrain a combination of the size and velocity.

In order to test whether this was possible we examined simulated
spectra.  1st and 2nd order \emph{XMM} RGS1 and 2 response matrices
were created with the \textsc{sas} \textsc{rgsrmfgen} tool. We
supplied a custom Gaussian angular distribution with width
$\sigma=0.3$~arcmin. Spectra were simulated with these responses in
\textsc{xspec} \citep{ArnaudXspec} using a 1.5 keV BVAPEC model,
$0.5\Zsun$ metallicity and including $500\kmps$ Doppler broadening. We
took these simulated spectra and fit them using a point-source
response matrix, a BVAPEC model, but convolved with a Gaussian with a
fixed width in wavelength (parameterised by an angular scale, as in
equation \ref{eqn:broad}). We were able to recover the broadening
velocity and Gaussian width of the source, $\sigma$, when fitting.

\subsection{Simple spatial modelling}
\label{sect:simplefit}

\begin{table*}
  \caption{Details of the examined objects and datasets. Listed are the
    object names, redshifts, extraction position in decimal degrees (J2000), mean
    cleaned RGS 1 and 2 exposure and \emph{XMM} datasets.}
 \begin{tabular}{lrrrrp{6cm}}
   Object & Redshift & RA & Dec & Exposure (ks) & \emph{XMM} observations \\ \hline
 2A 0335+096 & $0.0349$ & $54.6713$ & $9.9669$ & $117.1$ & 0147800201 \\
A1068 & $0.1375$ & $160.1855$ & $39.9529$ & $22.6$ & 0147630101 \\
A133 & $0.0566$ & $15.6740$ & $-21.8805$ & $24.0$ & 0144310101 \\
A1795 & $0.0625$ & $207.2182$ & $26.5937$ & $40.5$ & 0097820101 \\
A1835 & $0.2523$ & $210.2580$ & $2.8787$ & $220.6$ & 0098010101 0551830101 0551830201 \\
A1991 & $0.0587$ & $223.6318$ & $18.6449$ & $40.4$ & 0145020101 \\
A2029 & $0.0773$ & $227.7339$ & $5.7446$ & $172.1$ & 0111270201 0551780201 0551780301 0551780401 0551780501 \\
A2052 & $0.0355$ & $229.1855$ & $7.0214$ & $116.7$ & 0109920101 0401520301 0401520501 0401520601 0401520801 0401520901 0401521101 0401521201 0401521601 0401521701 \\
A2204 & $0.1522$ & $248.1957$ & $5.5753$ & $85.9$ & 0112230301 0306490101 0306490201 0306490301 0306490401 \\
A2597 & $0.0852$ & $351.3324$ & $-12.1243$ & $84.0$ & 0108460201 0147330101 \\
A262 & $0.0163$ & $28.1924$ & $36.1528$ & $127.7$ & 0109980101 0504780101 0504780201 \\
A2626 & $0.0553$ & $354.1272$ & $21.1467$ & $55.6$ & 0083150201 0148310101 \\
A3112 & $0.075252$ & $49.4902$ & $-44.2381$ & $203.3$ & 0105660101 0603050101 0603050201 \\
A3581 & $0.023$ & $211.8741$ & $-27.0183$ & $137.8$ & 0205990101 0504780301 0504780401 \\
A383 & $0.1871$ & $42.0140$ & $-3.5291$ & $32.3$ & 0084230501 \\
A4059 & $0.0475$ & $359.2542$ & $-34.7591$ & $46.6$ & 0109950101 0109950201 \\
A496 & $0.0329$ & $68.4077$ & $-13.2616$ & $149.5$ & 0135120201 0506260301 0506260401 \\
AS1101 & $0.058$ & $348.4948$ & $-42.7263$ & $98.0$ & 0147800101 \\
Centaurus & $0.0114$ & $192.2039$ & $-41.3125$ & $157.7$ & 0046340101 0406200101 \\
E 1455+2232 & $0.2578$ & $224.3129$ & $22.3424$ & $36.1$ & 0108670201 \\
HCG 62 & $0.0137$ & $193.2738$ & $-9.2042$ & $156.9$ & 0112270701 0504780501 0504780601 \\
Hercules A & $0.154$ & $252.7841$ & $4.9924$ & $119.8$ & 0401730101 0401730201 0401730301 \\
Hydra A & $0.0539$ & $139.5249$ & $-12.0955$ & $93.5$ & 0504260101 \\
MACS J0242.5-2132 & $0.314$ & $40.6497$ & $-21.5406$ & $69.0$ & 0673830101 \\
MACS J2229.7-2755 & $0.324$ & $337.4385$ & $-27.9270$ & $46.8$ & 0651240201 \\
MKW 3s & $0.045$ & $230.4662$ & $7.7088$ & $32.6$ & 0109930101 \\
MS 0735.6+7421 & $0.216$ & $115.4344$ & $74.2440$ & $55.2$ & 0303950101 \\
MS 2137.3-2353 & $0.313$ & $325.0632$ & $-23.6612$ & $63.9$ & 0673830201 \\
NGC 1316 & $0.00587$ & $50.6738$ & $-37.2079$ & $159.4$ & 0302780101 0502070201 \\
NGC 1365 & $0.005457$ & $53.4016$ & $-36.1416$ & $111.6$ & 0505140401 \\
NGC 1399 & $0.0046$ & $54.6212$ & $-35.4506$ & $126.0$ & 0400620101 \\
NGC 1404 & $0.00649$ & $54.7156$ & $-35.5946$ & $41.9$ & 0304940101 \\
NGC 1550 & $0.012389$ & $64.9080$ & $2.4100$ & $25.3$ & 0152150101 \\
NGC 2300 & $0.007$ & $113.0821$ & $85.7093$ & $52.5$ & 0022340201 \\
NGC 3411 & $0.0153$ & $162.6087$ & $-12.8451$ & $24.1$ & 0146510301 \\
NGC 4261 & $0.00706$ & $184.8468$ & $5.8249$ & $94.6$ & 0502120101 \\
NGC 4325 & $0.02571$ & $185.7776$ & $10.6210$ & $21.1$ & 0108860101 \\
NGC 4472 & $0.00332$ & $187.4449$ & $8.0008$ & $83.6$ & 0200130101 \\
NGC 4636 & $0.00313$ & $190.7080$ & $2.6876$ & $57.7$ & 0111190701 \\
NGC 4649 & $0.00373$ & $190.9165$ & $11.5527$ & $76.0$ & 0502160101 \\
NGC 499 & $0.01467$ & $20.7985$ & $33.4607$ & $49.1$ & 0501280101 \\
NGC 5044 & $0.00928$ & $198.8498$ & $-16.3854$ & $133.7$ & 0037950101 0554680101 \\
NGC 533 & $0.01739$ & $21.3808$ & $1.7592$ & $34.2$ & 0109860101 \\
NGC 5813 & $0.0066$ & $225.2969$ & $1.7020$ & $162.3$ & 0302460101 0554680201 0554680301 \\
NGC 5846 & $0.0057$ & $226.6223$ & $1.6049$ & $37.9$ & 0021540101 0021540501 \\
RBS 540 & $0.0397$ & $66.4636$ & $-8.5601$ & $41.1$ & 0300210401 \\
RBS 797 & $0.354$ & $146.8029$ & $76.3871$ & $17.7$ & 0502940301 \\
RXC J0605.8-3518 & $0.141$ & $91.4749$ & $-35.3026$ & $24.1$ & 0201901001 \\
RXC J1044.5-0704 & $0.1323$ & $161.1369$ & $-7.0691$ & $28.9$ & 0201901501 \\
RXC J1141.4-1216 & $0.1195$ & $175.3518$ & $-12.2777$ & $29.5$ & 0201901601 \\
RXC J1504.1-0248 & $0.2153$ & $226.0309$ & $-2.8044$ & $38.9$ & 0401040101 \\
RXC J2014.8-2430 & $0.1612$ & $303.7154$ & $-24.5057$ & $26.4$ & 0201902201 \\
RXC J2149.1-3041 & $0.1179$ & $327.2822$ & $-30.7013$ & $26.8$ & 0201902601 \\
RX J1720.1+2638 & $0.164$ & $260.0414$ & $26.6248$ & $69.2$ & 0500670201 0500670301 0500670401 \\
RX J2129.6+0005 & $0.235$ & $322.4164$ & $0.0886$ & $43.5$ & 0093030201 \\
Zw3146 & $0.2906$ & $155.9152$ & $4.1865$ & $209.6$ & 0108670101 0605540201 0605540301 \\

 \hline
 \end{tabular}
 \label{tab:objects}
\end{table*}

\begin{figure}
  \includegraphics[width=\columnwidth]{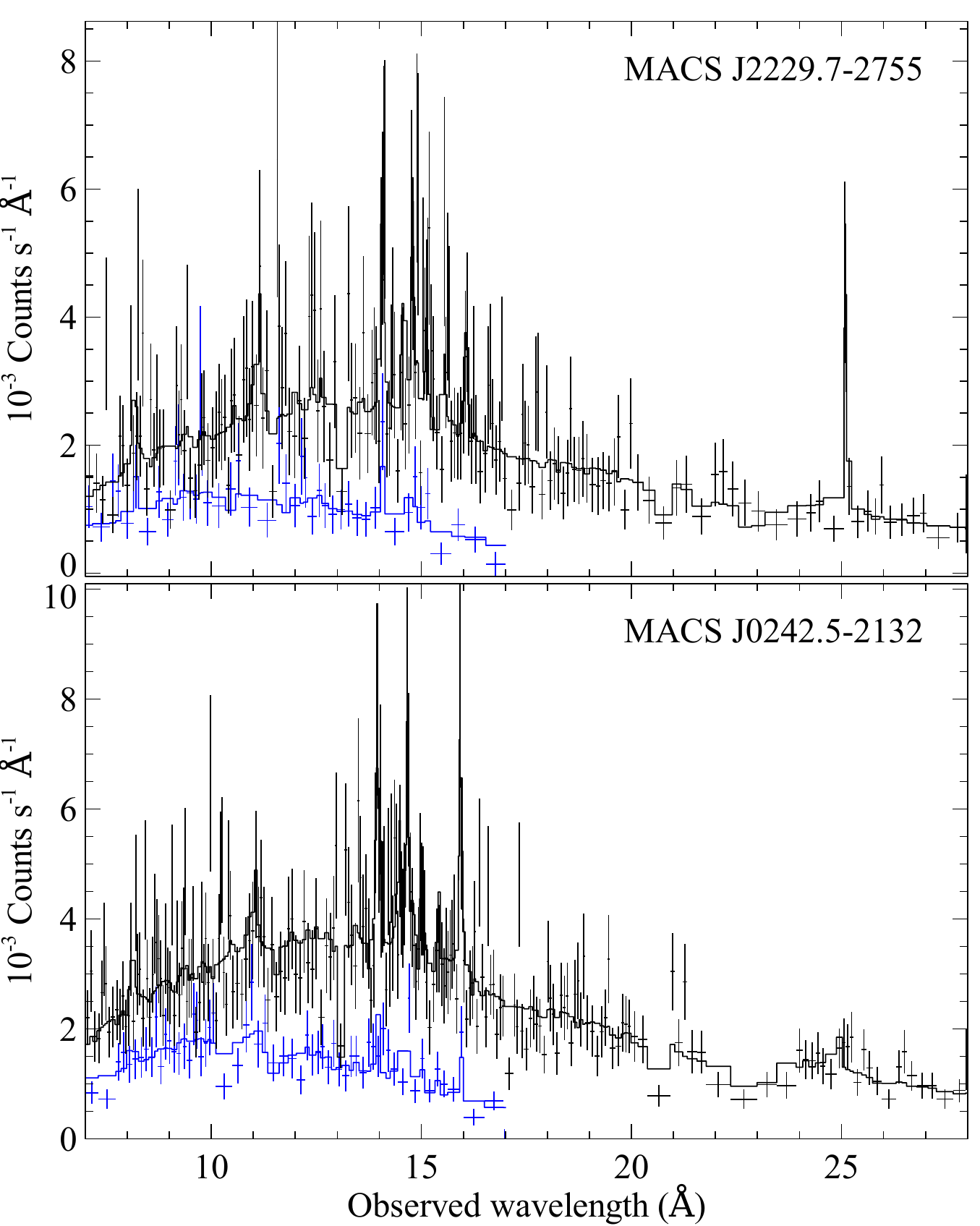}
  \caption{RGS spectra of two new observations of MACS clusters
    included in this paper. The data where rebinned for
    MACS\,J2229.7-2755 and MACS\,J0242.5-2132 to a minimum signal to
    noise ratio of 3 and 4, respectively. We note that
    MACS\,J0242.5-2132 does not show the obvious narrow
    O~\textsc{viii} emission line as seen in MACS\,J2229.7-2755. This
    appears to be due to a combination of a lower best-fitting oxygen
    metallicity ($0.15 \pm 0.08$ vs $0.44^{+0.18}_{-0.11}\Zsun$),
    broader emission lines and higher absorbing column density. The
    two clusters have similar temperatures, fluxes and redshifts.}
  \label{fig:newspec}
\end{figure}

The data were prepared using the methods described in
\cite{Sanders10_Broaden}. We used the same 90 per cent of the point
spread function (PSF) extraction region in the cross-dispersion
direction and 90 per cent of the pulse height distribution. This
selection region corresponds roughly to a 50 arcsec strip across the
cluster in the cross-dispersion direction. Background spectra were
extracted outside 98 per cent of the PSF. There have been some changes
to the datasets included. We list the datasets used in Table
\ref{tab:objects}. These data include observations of new objects such
as MACS\,J2229.7-2755 (see Fig.~\ref{fig:newspec}). We have removed
some of the poorer quality datasets from the sample, in particular
those where the background was a large fraction of the spectrum or the
emission lines were weak. The datasets were reprocessed to take
advantage of the updated calibration of the RGS line-spread function
(XMM-CCF-REL-275). In addition, we increased the number of rows in the
response matrices from the default value of 4000 to 16000.

For most of the objects we fitted a single component BVAPEC spectral
model (i.e. an APEC thermal model with variable abundances, thermal
broadening and variable Gaussian broadening). The APEC model
\citep{SmithApec01} we used was recreated from the APED 2.0.2 database
with a temperature grid size of 0.01 dex, rather than using the
default 0.1 dex tables. The PHABS photoelectric absorber component
\citep{BalucinskaChurchPhabs92} was applied to this model. We
convolved the absorbed thermal model with a Gaussian of a constant
wavelength width, parameterised as the spatial scale. The data were
fit in the \textsc{xspec} spectral analysis package, minimising the
C-statistic during fitting. We fit the spectra between 7 and {28\AA}
for the first order data and 7 to {17\AA} for the second order
spectra.

We note that using a spectral background extracted off-axis from the
dataset itself may not be correct and depends on the source morphology
and structure. If most of the flux from the object is emitted from
within the extraction region, the background components will be
correctly subtracted using off axis regions. If the source is more
extended than the background region, the background will contain
source emission. The extraction region is a strip across the cluster
and will contain emission from regions outside the core. Using an
adjacent strip as background (as we obtain from the off-axis data)
will remove much of this outer emission.  This spectral component, as
it comes from a more extended region will be more highly broadened
than that from the source region. We are only interested in the best
measurements from the centre, so it is favourable to remove the outer
emission. This will particularly the case for a cool core embedded in
a hotter medium. However, there are cases where the geometry of the
source may make this subtraction invalid, particularly if the source
is not symmetric in the extraction regions.  For some objects it would
be better to use background spectra from blank sky observations. We
also redo the fits using template backgrounds generated with the
\textsc{rgsbkgmodel} tool to check the robustness of our backgrounds.

\begin{table}
  \caption{Spectral fitting parameters used when fitting
    spectra. $N_T$ is the number of temperature components used when
    fitting the spectra and $N_S$ is the number of spatial
    components. If two spatial and temperature components are listed,
    two spatial components with independent temperatures
    were fitted. The listed metallicities were allowed to vary
    independently. If Fe is not listed, it was frozen at the Solar
    value because of a lack of continuum. Unlisted metals were tied to
    the Fe Solar ratio, except for He which was frozen at the Solar
    ratio. In addition to these parameters, we allowed the absorbing
    column density, normalisation (for each temperature component or
    spatial component), redshift and velocity width to vary.}
  \begin{tabular}{llll}
    Object & $N_T$ & $N_S$ & Metallicities \\ \hline
    2A 0335+096 & 1 & 1 & O Ne Mg Si Fe Ni \\
A1068 & 1 & 1 & O Ne Mg Si Fe Ni \\
A133 & 1 & 1 & O Ne Mg Fe Ni \\
A1795 & 1 & 1 & O Ne Mg Si Fe Ni \\
A1835 & 1 & 1 & O Ne Mg Si Fe Ni \\
A1991 & 1 & 1 & O Ne Mg Si Fe Ni \\
A2029 & 1 & 1 & O Ne Mg Si Fe Ni \\
A2052 & 1 & 1 & O Ne Mg Fe Ni \\
A2204 & 1 & 1 & O Ne Mg Si Fe Ni \\
A2597 & 1 & 1 & O Ne Mg Fe Ni \\
A262 & 2 & 2 & O Ne Mg Si Fe Ni \\
A2626 & 1 & 1 & O Ne Mg Si Fe Ni \\
A3112 & 1 & 1 & O Ne Mg Si Fe Ni \\
A3581 & 2 & 2 & O Ne Mg Fe Ni \\
A383 & 1 & 1 & O Ne Mg Si Fe Ni \\
A4059 & 1 & 1 & O Ne Mg Fe Ni \\
A496 & 1 & 1 & O Ne Mg Fe Ni \\
AS1101 & 1 & 1 & O Ne Mg Si Fe Ni \\
Centaurus & 2 & 2 & O Ne Mg Fe Ni \\
E 1455+2232 & 1 & 1 & O Mg Si Fe \\
HCG 62 & 2 & 2 & N O Ne Mg Fe Ni \\
Hercules A & 1 & 1 & O Ne Mg Si Fe Ni \\
Hydra A & 1 & 2 & O Ne Mg Fe Ni \\
MACS J0242.5-2132 & 1 & 1 & O Ne Mg Si Fe Ni \\
MACS J2229.7-2755 & 1 & 1 & O Ne Mg Si Fe Ni \\
MKW 3s & 1 & 1 & O Ne Mg Si Fe Ni \\
MS 0735.6+7421 & 1 & 1 & O Ne Mg Si Fe Ni \\
MS 2137.3-2353 & 1 & 1 & O Ne Mg Si Fe Ni \\
NGC 1316 & 1 & 1 & O Ne Mg Fe Ni \\
NGC 1365 & 1 & 1 & O Ne Mg Fe Ni \\
NGC 1399 & 1 & 1 & N O Ne Mg Fe Ni \\
NGC 1404 & 1 & 1 & O Ne Mg Fe Ni \\
NGC 1550 & 1 & 1 & O Ne Mg Fe Ni \\
NGC 2300 & 1 & 1 & O Ne Mg Ni \\
NGC 3411 & 1 & 1 & O Ne Mg Fe Ni \\
NGC 4261 & 1 & 1 & O Ne Mg Fe Ni \\
NGC 4325 & 1 & 1 & O Ne Mg Ni \\
NGC 4472 & 1 & 1 & O Ne Mg Fe Ni \\
NGC 4636 & 1 & 1 & O Ne Mg Fe Ni \\
NGC 4649 & 1 & 2 & O Ne Mg Fe Ni \\
NGC 499 & 1 & 1 & O Ne Mg Fe Ni \\
NGC 5044 & 1 & 2 & O Ne Mg Fe Ni \\
NGC 533 & 1 & 1 & O Ne Mg Ni \\
NGC 5813 & 2 & 2 & N O Ne Mg Fe Ni \\
NGC 5846 & 1 & 1 & O Ne Mg Ni \\
RBS 540 & 1 & 1 & O Ne Mg Si Fe Ni \\
RBS 797 & 1 & 1 & O Ne Mg Fe \\
RXC J0605.8-3518 & 1 & 1 & O Ne Mg Si Fe Ni \\
RXC J1044.5-0704 & 1 & 1 & O Ne Mg Si Fe Ni \\
RXC J1141.4-1216 & 1 & 1 & O Ne Mg Fe \\
RXC J1504.1-0248 & 1 & 1 & O Ne Mg Si Fe Ni \\
RXC J2014.8-2430 & 1 & 1 & O Ne Mg Si Fe \\
RXC J2149.1-3041 & 1 & 1 & O Ne Mg Si Fe \\
RX J1720.1+2638 & 1 & 1 & O Ne Mg Si Fe Ni \\
RX J2129.6+0005 & 1 & 1 & O Mg Si Fe Ni \\
Zw3146 & 1 & 1 & O Ne Mg Si Fe Ni \\

    \hline
  \end{tabular}
  \label{tab:params}
\end{table}

We allowed the temperature, absorption, redshift, velocity broadening,
spatial broadening, normalisation and some metallicities to be free in
the spectral fits. The metals which we allowed to be free depended on
the temperature of the object and whether there was any continuum in
the spectrum. For some objects with high quality data or when required
we fitted more than one broadened thermal component, to allow for more
than one temperature or spatial scale. The parameters which were
allowed to be free during the spectral fitting for each object are
shown in Table \ref{tab:params}.

We examined the uncertainties in the parameters to the fits using both
a conventional approach of varying the parameter until the fit
statistic increases from the minimum by the required amount (referred
to here as $\delta$C) and using Markov Chain Monte Carlo
(MCMC). Rather than use the standard Metropolis-Hastings MCMC
algorithm, we used an affine-invariant MCMC sampler
\citep{GoodmanWeare10} as implemented in \textsc{emcee}\footnote{See
  \url{http://danfm.ca/emcee/}} \citep{ForemanMackey12}. We created a
package based on \textsc{emcee} to analyse X-ray spectral models and
data in parallel invocations of \textsc{xspec}\footnote{Available at
  \url{https://github.com/jeremysanders/xspec_emcee}}. The advantage
of an affine-invariant sampler over the Metropolis-Hastings algorithm
is that it can sample complex parameter space much more efficiently
without any tuning of a proposal distribution. It uses a number of
`walkers' which simultaneously examine parameter space within each
step on the chain. The proposal distribution for a particular walker
depends on the position of the other walkers.

\begin{table*}
  \caption{Velocity broadening results from the change
    in fit statistic ($\delta$C),
    both with standard off-axis and template backgrounds, and
    from the MCMC analysis. We show 1$\sigma$ and 90 per cent confidence
    uncertainties ($\kmps$). The number of walkers used in the MCMC analysis is
    listed. We also show the temperature from the MCMC analysis, with
    $1\sigma$ uncertainties. If more than one temperature component
    was used, this is the value of the higher component. A blank MCMC
    result indicates that the limit was greater than $5000\kmps$.}
  \begin{tabular}{lcccccccc}
    Object & $\delta$C   & $\delta$C    &  $\delta$C Template  & $\delta$C Template   & MCMC       & MCMC         & MCMC     & Temperature \\
           & ($1\sigma$) & (90 per cent)& ($1\sigma$)  & (90 per cent)& ($1\sigma$)& (90 per cent)& walkers  & (keV) \\
    \hline
    2A 0335+096 & $<190$ & $<300$ & $<210$ & $<330$ & $<200$ & $<300$ & $200$ & $1.58 \pm 0.02$ \\
A1068 & $<490$ & $<710$ & $<350$ & $<570$ & $<480$ & $<760$ & $200$ & $2.89^{+0.43}_{-0.38}$ \\
A133 & $<420$ & $<580$ & $490^{+160}_{-150}$ & $490^{+280}_{-250}$ & $<380$ & $<540$ & $200$ & $3.12^{+0.59}_{-0.42}$ \\
A1795 & $<550$ & $<800$ & $<500$ & $<750$ & $<580$ & $<880$ & $200$ & $3.35^{+0.41}_{-0.32}$ \\
A1835 & $<240$ & $<310$ & $<220$ & $<290$ & $<200$ & $<300$ & $200$ & $4.58^{+0.67}_{-0.49}$ \\
A1991 & $<390$ & $<460$ & $<380$ & $<490$ & $210^{+110}_{-170}$ & $<400$ & $200$ & $1.68 \pm 0.06$ \\
A2029 & $<410$ & $<480$ & $<470$ & $<530$ & $270^{+70}_{-230}$ & $<420$ & $200$ & $3.82^{+0.44}_{-0.29}$ \\
A2052 & $<550$ & $<700$ & $400^{+220}_{-340}$ & $400 \pm 370$ & $<440$ & $<640$ & $200$ & $1.72 \pm 0.03$ \\
A2204 & $<440$ & $<520$ & $330^{+110}_{-230}$ & $<510$ & $310^{+130}_{-190}$ & $<480$ & $200$ & $3.68^{+0.46}_{-0.33}$ \\
A2597 & $460^{+190}_{-430}$ & $<750$ & $<580$ & $<680$ & $550^{+130}_{-310}$ & $550^{+170}_{-510}$ & $200$ & $2.92^{+0.25}_{-0.21}$ \\
A2626 & $<470$ & $<670$ & $<690$ & $<900$ & $<440$ & $<660$ & $4000$ & $3.49^{+0.88}_{-0.57}$ \\
A262$^{2}$ & $<400$ & $<570$ & $<590$ & $<680$ & $<400$ & $<560$ & $200$ & $1.41^{+0.04}_{-0.04}$ \\
A3112 & $420^{+100}_{-140}$ & $420^{+150}_{-270}$ & $440^{+100}_{-150}$ & $440^{+160}_{-310}$ & $430^{+110}_{-130}$ & $430^{+150}_{-270}$ & $200$ & $3.11^{+0.13}_{-0.12}$ \\
A3581$^{2}$ & $<300$ & $<460$ & $<400$ & $<590$ & $<280$ & $<420$ & $200$ & $1.41^{+0.03}_{-0.02}$ \\
A383 & $<450$ & $<690$ & $<480$ & $<740$ & $<480$ & $<780$ & $4000$ & $3.61^{+0.85}_{-0.65}$ \\
A4059 & $940^{+370}_{-690}$ & $<1520$ & $940^{+350}_{-500}$ & $<1470$ & $1050^{+390}_{-550}$ & $<1540$ & $4000$ & $2.69^{+0.43}_{-0.29}$ \\
A496 & $<550$ & $<690$ & $540^{+200}_{-330}$ & $<850$ & $<460$ & $<640$ & $200$ & $2.19^{+0.06}_{-0.06}$ \\
AS1101 & $<290$ & $<440$ & $<310$ & $<420$ & $<280$ & $<440$ & $200$ & $2.24^{+0.08}_{-0.07}$ \\
Centaurus$^{2}$ & $<250$ & $<340$ & $<370$ & $<450$ & $<240$ & $<340$ & $200$ & $1.57 \pm 0.02$ \\
E 1455+2232 & $<180$ & $<300$ & $<220$ & $<350$ & $<260$ & $<540$ & $4000$ & $3.36^{+1.11}_{-0.56}$ \\
HCG 62$^{2}$ & $390^{+110}_{-120}$ & $390^{+170}_{-220}$ & $460 \pm 80$ & $460 \pm 140$ & $390^{+130}_{-150}$ & $390^{+190}_{-290}$ & $200$ & $0.93^{+0.02}_{-0.02}$ \\
Hercules A & $<810$ & $<1070$ & $<770$ & $<1050$ & $<800$ & $<1060$ & $4000$ & $4.83^{+1.27}_{-1.16}$ \\
Hydra A$^{2}$ & $500^{+160}_{-230}$ & $<810$ & $480^{+200}_{-210}$ & $<800$ & $490^{+210}_{-270}$ & $490^{+290}_{-450}$ & $200$ & $3.09^{+0.26}_{-0.22}$ \\
MACS J0242.5-2132 & $<450$ & $<560$ & $<470$ & $<570$ & $330^{+170}_{-290}$ & $<600$ & $200$ & $4.89^{+1.04}_{-0.98}$ \\
MACS J2229.7-2755 & $<150$ & $<240$ & $<180$ & $<280$ & $<180$ & $<280$ & $200$ & $3.80^{+0.73}_{-0.63}$ \\
MKW 3s & $<560$ & $<870$ & $<970$ & $<1480$ & $<760$ & $<1220$ & $4000$ & $3.04^{+0.37}_{-0.27}$ \\
MS 0735.6+7421 & $<370$ & $<540$ & $<200$ & $<360$ & $<780$ & $<1360$ & $4000$ & $4.46^{+1.38}_{-1.13}$ \\
MS 2137.3-2353 & $460 \pm 200$ & $460^{+330}_{-370}$ & $420^{+210}_{-220}$ & $<750$ & $390^{+190}_{-370}$ & $<800$ & $200$ & $6.18^{+1.17}_{-1.08}$ \\
NGC 1316 & $<250$ & $<420$ & $<460$ & $<560$ & $<320$ & $<440$ & $200$ & $0.64 \pm 0.01$ \\
NGC 1365 & $<350$ & $<460$ & $<440$ & $<490$ & $<340$ & $<420$ & $200$ & $0.54^{+0.04}_{-0.05}$ \\
NGC 1399 & $380^{+90}_{-130}$ & $380^{+130}_{-260}$ & $520^{+50}_{-60}$ & $520^{+90}_{-90}$ & $390^{+90}_{-110}$ & $390^{+130}_{-250}$ & $200$ & $0.88^{+0.01}_{-0.01}$ \\
NGC 1404 & $<330$ & $<530$ & $<340$ & $<520$ & $<360$ & $<600$ & $200$ & $0.64 \pm 0.01$ \\
NGC 1550 & $<990$ & $<1550$ & $<1120$ & $<1620$ & $<980$ & $<1760$ & $4000$ & $1.12^{+0.02}_{-0.03}$ \\
NGC 2300 & $<240$ & $<390$ & $<360$ & $<590$ & $<280$ & $<480$ & $200$ & $0.69 \pm 0.02$ \\
NGC 3411 & $<630$ & $<950$ & $<770$ & $<980$ & $<700$ & $<1100$ & $200$ & $0.89^{+0.02}_{-0.02}$ \\
NGC 4261 & $<230$ & $<330$ & $<350$ & $<390$ & $<200$ & $<300$ & $200$ & $0.71 \pm 0.01$ \\
NGC 4325 & $<600$ & $<760$ & $<380$ & $<570$ & $<620$ & $<700$ & $200$ & $0.86^{+0.01}_{-0.02}$ \\
NGC 4472 & $<140$ & $<220$ & $<150$ & $<240$ & $<140$ & $<240$ & $200$ & $0.82 \pm 0.01$ \\
NGC 4636 & $<450$ & $<550$ & $<480$ & $<580$ & $270^{+130}_{-230}$ & $<500$ & $200$ & $0.63^{+0.01}_{-0.01}$ \\
NGC 4649$^{2}$ & $<320$ & $<400$ & $<280$ & $<360$ & $<280$ & $<400$ & $200$ & $0.85 \pm 0.01$ \\
NGC 499 & $<1190$ & $<1850$ & $<2210$ & $<2630$ & $<1240$ & $<1940$ & $4000$ & $0.73 \pm 0.02$ \\
NGC 5044$^{2}$ & $<470$ & $<570$ & $400^{+150}_{-300}$ & $<630$ & $370^{+110}_{-270}$ & $<540$ & $200$ & $0.82 \pm 0.01$ \\
NGC 533 & $<230$ & $<380$ & $<360$ & $<530$ & $<280$ & $<420$ & $200$ & $0.81 \pm 0.02$ \\
NGC 5813$^{2}$ & $360^{+180}_{-280}$ & $<620$ & $450^{+160}_{-200}$ & $<680$ & $430^{+130}_{-230}$ & $<580$ & $200$ & $0.73^{+0.01}_{-0.01}$ \\
NGC 5846 & $<330$ & $<480$ & $<500$ & $<630$ & $<320$ & $<480$ & $200$ & $0.66^{+0.02}_{-0.01}$ \\
RBS 540 & $390^{+150}_{-160}$ & $390^{+250}_{-340}$ & $570^{+150}_{-130}$ & $570^{+260}_{-220}$ & $<500$ & $<580$ & $200$ & $2.52^{+0.23}_{-0.18}$ \\
RBS 797 & $<340$ & $<540$ & $<410$ & $<720$ & $<2500$ &  & $4000$ & $5.24^{+1.57}_{-1.27}$ \\
RX J1720.1+2638 & $<380$ & $<730$ & $<480$ & $<820$ & $<600$ & $<1040$ & $4000$ & $5.36^{+1.29}_{-1.05}$ \\
RX J2129.6+0005 & $<270$ & $<1780$ & $<460$ & $<1600$ & $<600$ & $<1580$ & $200$ & $5.00^{+1.18}_{-1.04}$ \\
RXC J0605.8-3518 & $<300$ & $<490$ & $<340$ & $<520$ & $<380$ & $<640$ & $4000$ & $6.11^{+1.76}_{-1.61}$ \\
RXC J1044.5-0704 & $20 \pm 490$ & $<730$ & $<530$ & $<710$ & $<500$ & $<880$ & $4000$ & $3.43^{+0.71}_{-0.56}$ \\
RXC J1141.4-1216 & $<280$ & $<480$ & $<300$ & $<490$ & $<360$ & $<600$ & $4000$ & $2.04^{+0.22}_{-0.16}$ \\
RXC J1504.1-0248 & $670^{+600}_{-360}$ & $670^{+950}_{-630}$ & $800^{+480}_{-530}$ & $<1570$ & $1310^{+570}_{-670}$ & $1310^{+970}_{-1090}$ & $4000$ & $6.22^{+1.58}_{-1.23}$ \\
RXC J2014.8-2430 & $<200$ & $<420$ & $<330$ & $<750$ & $<480$ & $<820$ & $4000$ & $7.86 \pm 1.43$ \\
RXC J2149.1-3041 & $<200$ & $<310$ & $<230$ & $<350$ & $<220$ & $<360$ & $200$ & $2.67^{+0.95}_{-0.39}$ \\
Zw3146 & $310^{+90}_{-100}$ & $310^{+150}_{-170}$ & $310 \pm 90$ & $310 \pm 150$ & $290^{+70}_{-250}$ & $<420$ & $200$ & $3.94^{+0.54}_{-0.44}$ \\

    \hline
  \end{tabular}
  \label{tab:mainresults}
\end{table*}

We started the walkers in a small cluster around the best fitting
parameter values. 200 walkers were typically used, with a chain length
of 2000 (per walker) following a burn-in period of 400. Repeat
fractions of the chains after burn-in ranged between 65 and 77 per
cent. For some objects, we found discrepancies between the MCMC and
$\delta$C statistic approaches or there were problems in the
convergence of the chain. In these cases we used 4000 walkers, but
decreased the length of the chain to 1000 and the burn-in period to
200 to finish the analysis in a reasonable time period (Table
\ref{tab:mainresults} lists the number of walkers). When analysing the
posterior probability distributions we use all of the walkers. The
mean autocorrelation time of the velocity parameter for the walkers
are typically less than 3 per cent of the chain length and at worst 6
per cent of the chain length. We also examined the chains by eye to
look for problems in the convergence.

In Table \ref{tab:mainresults} are presented the broadening results
for each of the objects. The first two columns show the 1$\sigma$ and
90 per cent uncertainties (or limits) on the line-of-sight velocity
obtained by allowing the fit statistic to increase by 1.0 or 2.71,
respectively. These are followed by the same results when template
background spectra are used. The next two columns show the same
results obtained from the MCMC analysis. It is unclear how to
calculate an error bar or limit from MCMC when a parameter is close to
a hard limit ($0\kmps$ broadening here), relative to its
uncertainty. Normally one would take the median value of a parameter
and its percentiles from the chain. If the posterior probability
distribution is significant at a limit then using a percentile to
compute uncertainties will give finite values, rather than a value
consistent with that limit. As many of our velocities are limits, we
therefore used a different technique examining the marginalised
posterior probability distribution. This distribution was computed for
the velocity broadening in 250 linear velocity bins up to
$5000\kmps$. We selected the most likely velocity (this is the quoted
value when significant). We moved down to lower values of marginalised
probability until the required integrated probability was
obtained. The error bar that we quote are the minimum and maximum
values of velocity that bound this integrated probability region.

\begin{figure}
  \includegraphics[width=\columnwidth]{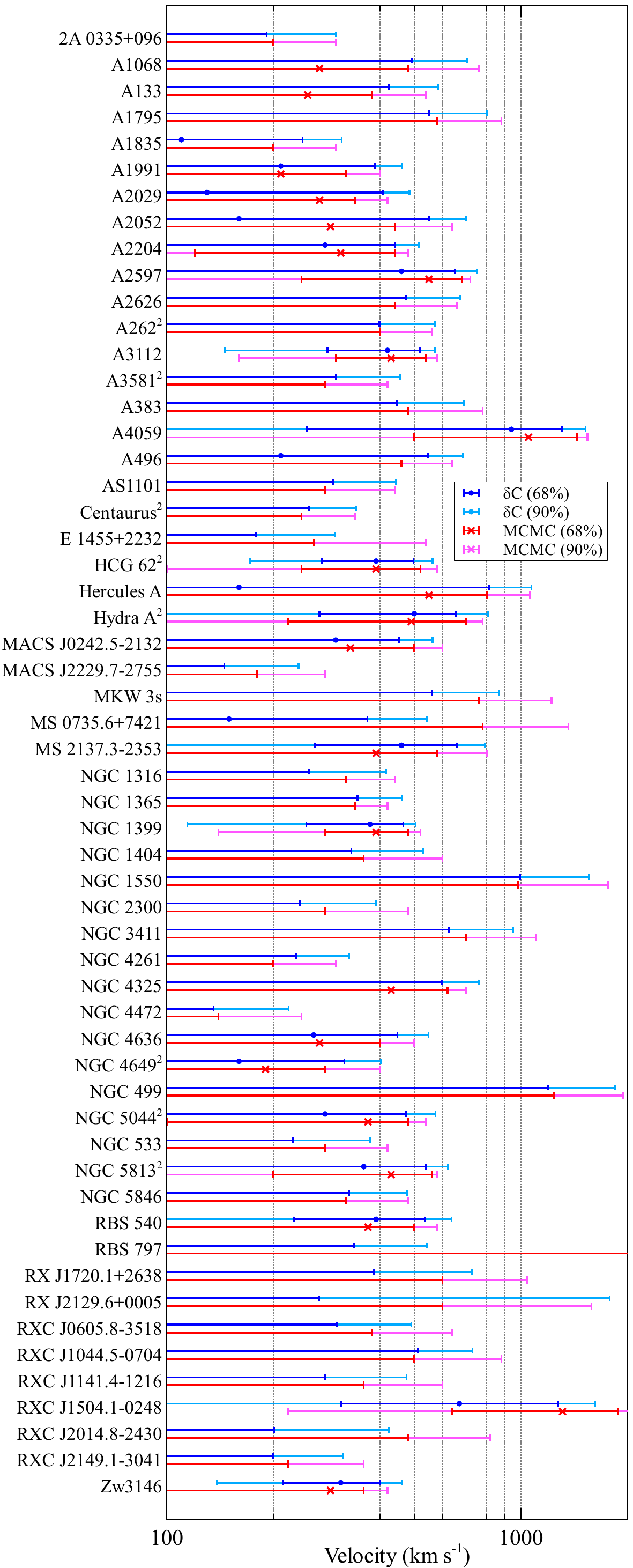}
  \caption{Error bars showing the $1\sigma$ (68 per cent) and 90 per
    cent confidence regions for each object. Results using $\delta$C
    statistic and MCMC posterior probabilities are shown for each
    object. Objects marked with $^2$ include two spatial and/or
    thermal components in the spectral fit.}
  \label{fig:sample}
\end{figure}

We plot our values and uncertainties from the standard and MCMC
methods for each object in Fig.~\ref{fig:sample}. It can be seen that
for most of the targets that we obtain good agreement between the two
methods. However, there are a few cases in which they do not show
consistent results. These include RBS\,797, where examination of the
MCMC chain shows a long tail on the velocity posterior distribution to
high velocities and a secondary minimum in the fit statistic at high
velocities. MS\,0735.6+7421 shows a similar tail to high velocities,
but the limit we can obtain is tighter. The tail on E\,1455+2232 and
RXC\,J2014.8-2430 are also longer than expected from the $\delta$C
analysis. The two methods agree for RXC\,J1504.1-0248 at the $1\sigma$
level. Some objects show rather poor constraints (e.g. NGC\,499).
Using template backgrounds makes little difference to the results for
most objects, although the change can be $\sim 200 \kmps$. The worst
cases include NGC\,499 and MKW\,3s, where the discrepancies are
$600-800 \kmps$.

\begin{figure}
  \includegraphics[width=\columnwidth]{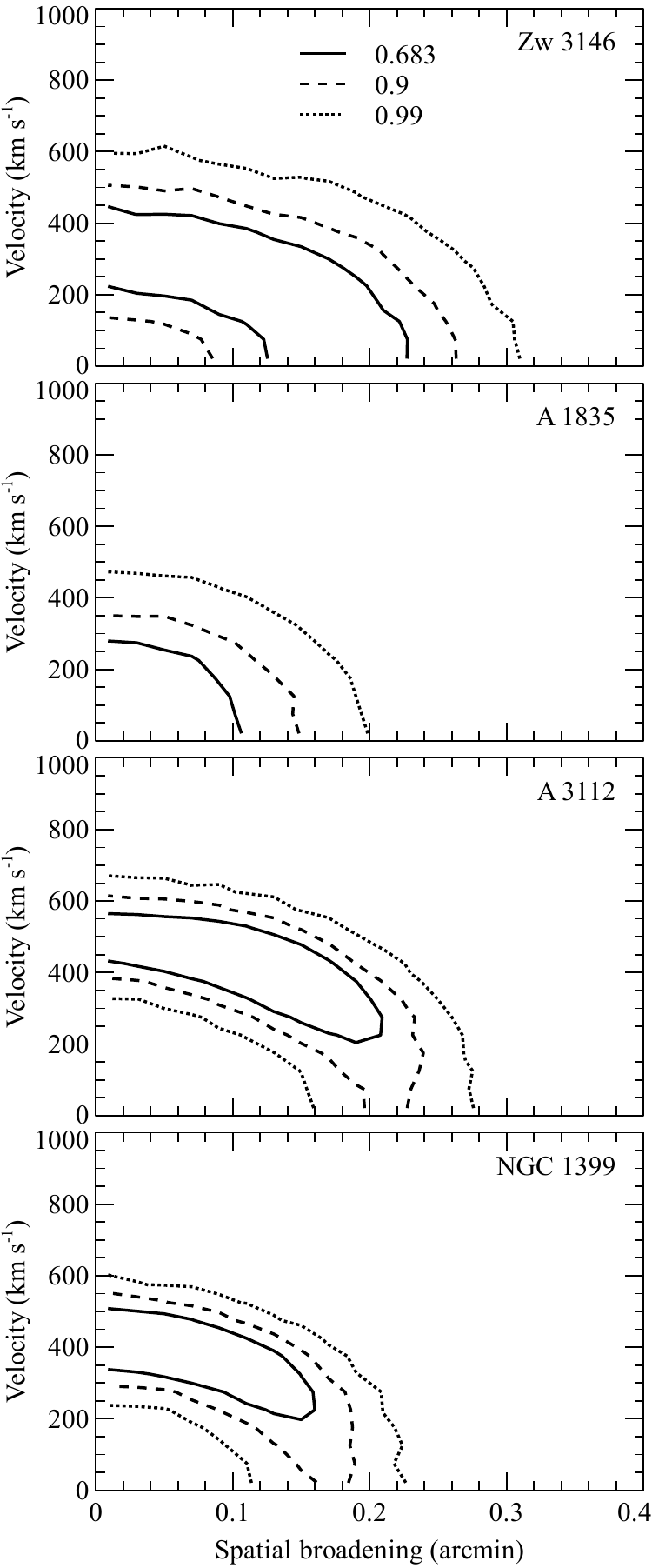}
  \caption{Confidence contours of velocity versus spatial broadening
    from the MCMC analysis. Contours show the 68.3, 90 and 95 per cent
    posterior confidence probabilities.}
  \label{fig:vel_size_conf}
\end{figure}

In the determination of the broadening of the emission lines the main
uncertainty is the extent of the source. We show in
Fig.~\ref{fig:vel_size_conf} the MCMC posterior probability contours
of the velocity and spatial broadening for four objects. It can be
seen that there is, of course, some degeneracy between the velocity
and spatial broadening. However, this is not complete, as can be seen
in A\,3112 and NGC\,1399.

\begin{figure}
  \includegraphics[width=\columnwidth]{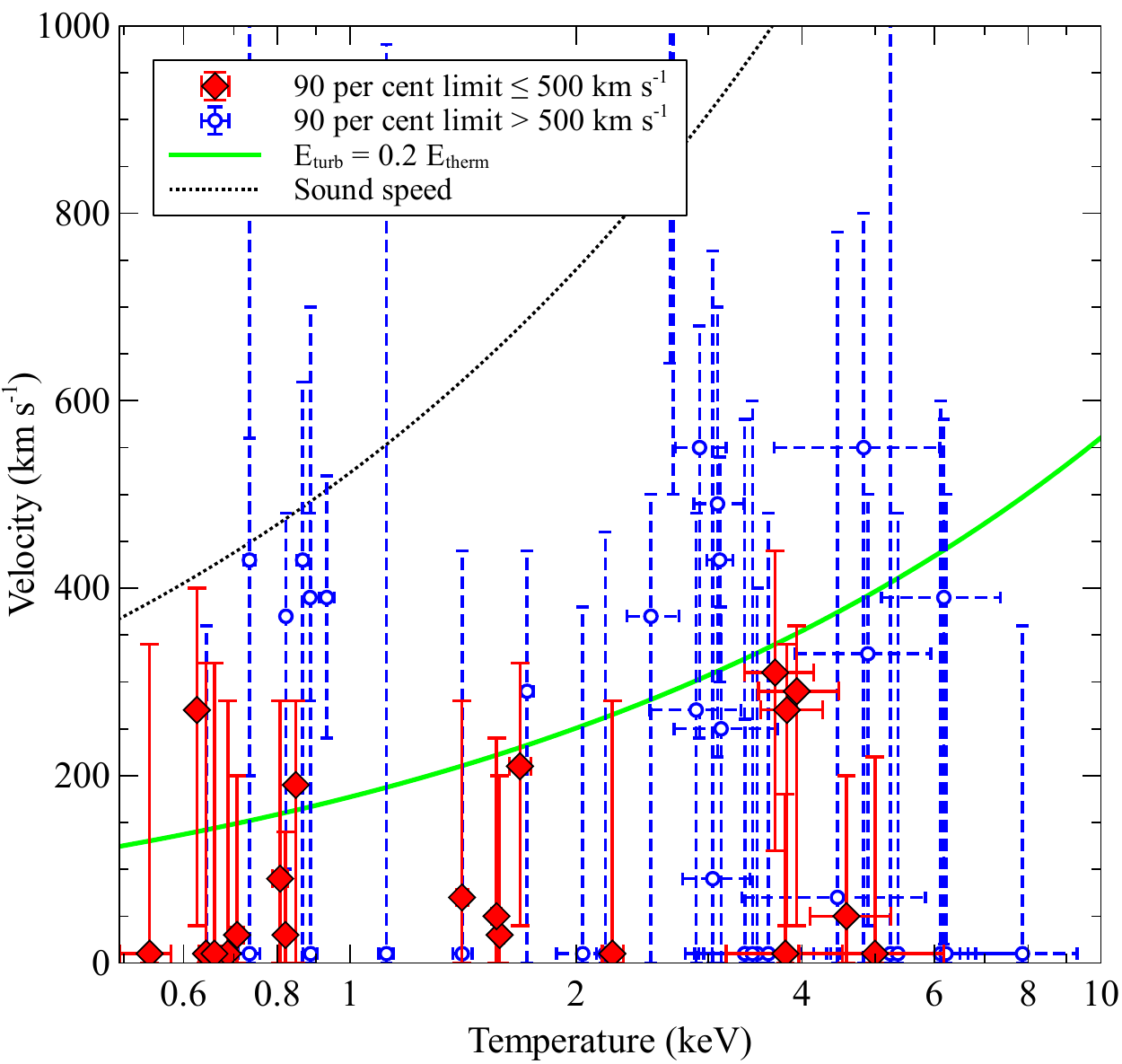}
  \caption{RGS temperature plotted against line width (using $1\sigma$
    MCMC uncertainties). We plot the upper temperature if more than
    one temperature component is fitted. The lower curve shows where
    the energy density in random motions is 20 per cent of the thermal
    energy density. The upper curve shows the sound speed.}
  \label{fig:T_vs_vel}
\end{figure}

In Fig.~\ref{fig:T_vs_vel} we plot the temperature of each object
against the best-fitting velocity ($1\sigma$ uncertainties,
determining the quantities using MCMC).

\subsection{Incorporation of \emph{Chandra} imaging}
\label{sect:imaging}
The most obvious improvement to our method of fitting for the spatial
extent of the source, is to use an X-ray image of the object from
\emph{Chandra} or \emph{XMM} to measure the spatial broadening which
should be included when modelling the spectrum. The profile of the
image along the RGS dispersion direction is examined to compute how
the spectrum is broadened. Either a wide band or narrower Fe-L band
image could be used. X-ray CCD images have poor spectral resolution,
so truly narrow band images cannot be made. Previously we used the
RGSXSRC model in \textsc{xspec} to account for the spatial broadening
in A\,1835 \citep{Sanders10_A1835}. \cite{Bulbul12} used MOS images
and \textsc{rgsrmfsmooth} to account for source extent in A\,3112.

\subsubsection{Examining A\,3112}

\begin{figure}
  \includegraphics[width=\columnwidth]{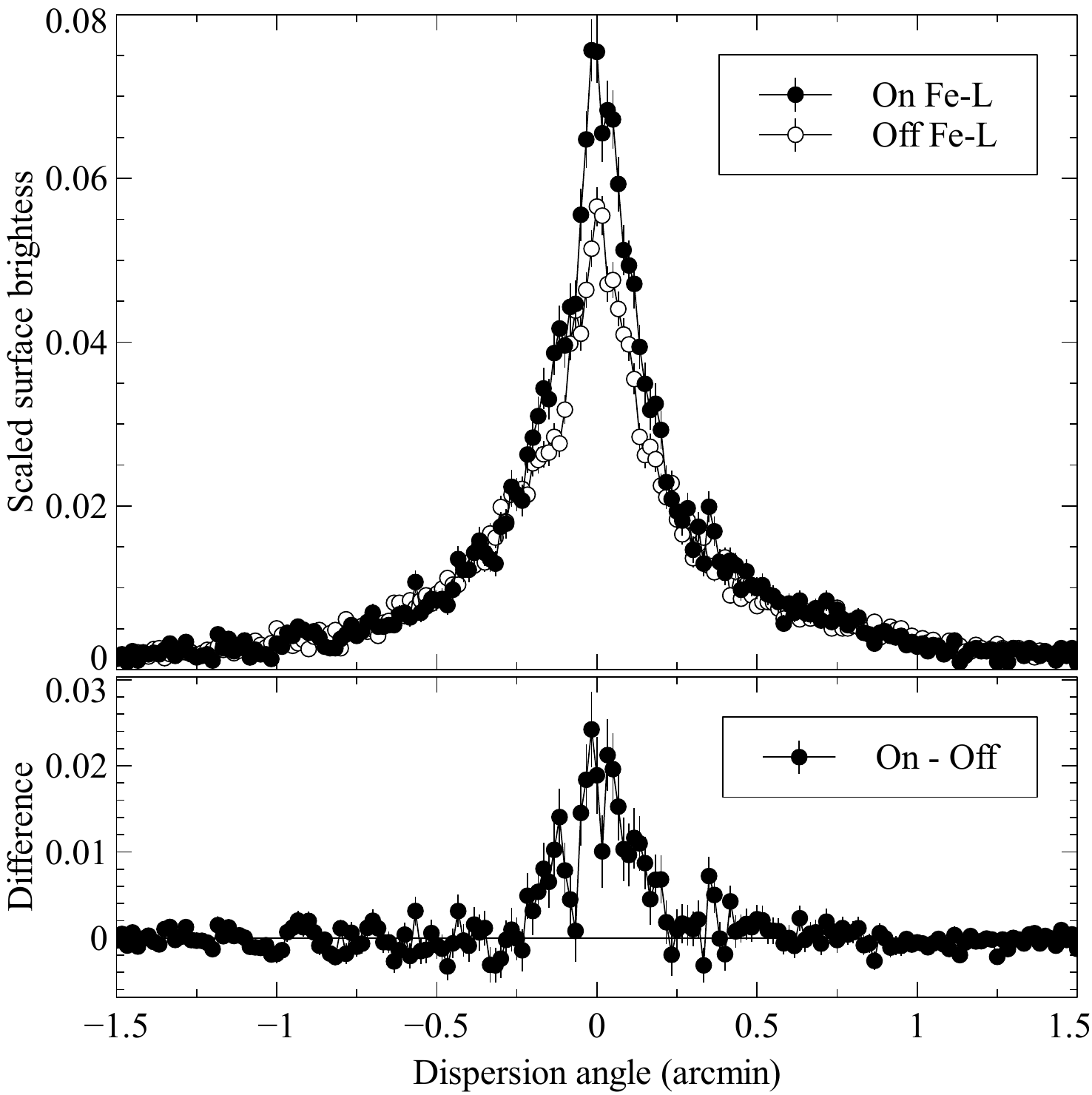}
  \caption{(Upper panel) \emph{Chandra} surface brightness profile
    along the dispersion direction for the \emph{XMM} 0603050101
    observation for A\,3112. The profiles are shown in the rest frame
    Fe-L (B-band, `On') and in the spectral region either side (A and
    C, `Off') and are normalized at angles of 1.37 to 1.9
    arcmin. (Lower panel) Continuum-subtracted Fe-L emission, computed
    from difference between the B-band Fe-L and A+B profiles.}
  \label{fig:sbprofile}
\end{figure}

\begin{figure}
  \includegraphics[width=\columnwidth]{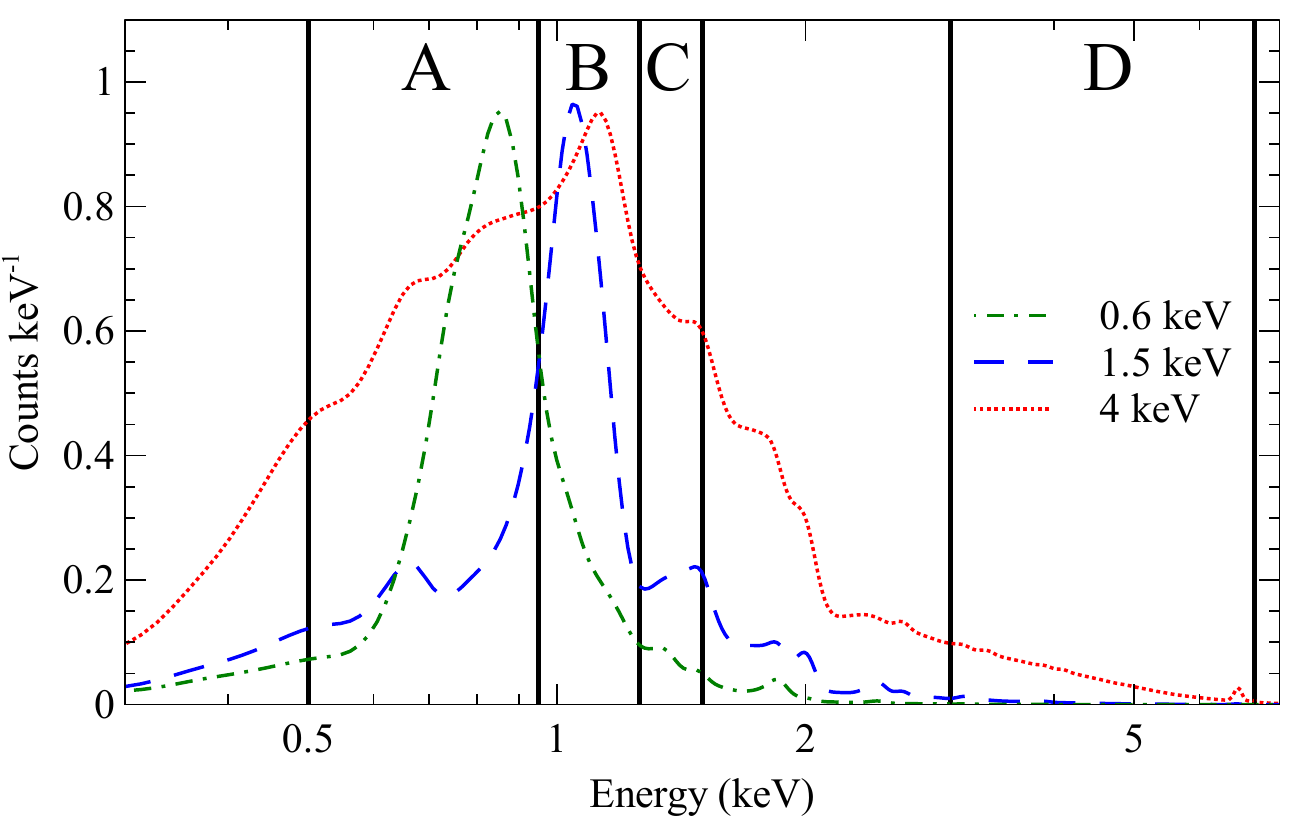}
  \caption{\emph{Chandra} bands used for making line images for RGS
    analysis. Model spectra at example temperatures and $0.5\Zsun$ are
    plotted. The effect of redshift were included.}
  \label{fig:bands}
\end{figure}

Including a surface brightness profile in the modelling will fail if
this profile does not accurately follow the emission lines that the
RGS spectral fits are sensitive to. It might be thought that choosing
a narrow band around Fe-L would give a good characterization of the
line surface brightness. However, our testing has shown that even
narrow bands often include too much continuum for many cool core
clusters at larger radii, leading to over-correction of the line
widths. The continuum at larger radii creates large wings in the
response matrix which control the spectral fits. We demonstrate this
in Fig.~\ref{fig:sbprofile} which shows \emph{Chandra} surface
brightness profiles along the dispersion angle for an RGS observation
of A\,3112. The profiles are shown in a band tuned to the peak of the
Fe-L emission (band B in Fig.~\ref{fig:bands}) and the sum of two
bands either side (bands A and C). The profiles are made in 1 arcsec
bins along the dispersion angle, selecting the cross dispersion angle
between $-0.45$ and $0.45$ arcmin. They are scaled by the number of
counts in each band in an annulus from 1.4 and 1.9 arcmin radius.

The surface brightness profiles continue to decline smoothly to large
radii in both the Fe-L and continuum-like bands. If the profiles in
the bands are subtracted (Fig.~\ref{fig:sbprofile} bottom panel), the
continuum-subtracted Fe-L emission is seen to be more compact, without
the long tails at large radii. The strong Fe-L emitting region
corresponds roughly to where the temperature drops below 3 keV in the
cluster.

We used the Fe-L `On' (B band) profile to create response matrices
using the `angdistset' option to the \textsc{rgsrmfgen} response
generation tool. This option broadens the response matrix to account
for the broadening due to the dispersion axis profile given. We input
the profiles between dispersion angles between $-4$ and $4$
arcmin. This was done for each of the A\,3112 datasets 0105660101,
0603050101 and 0603050201, as the dispersion angles were different in
each case. The spectra and response matrices were combined with
\textsc{rgscombine} and fitted in \textsc{xspec}. In this case we
obtained a best fitting broadening of $0 \kmps$ and an upper limit of
$175\kmps$ (90 per cent confidence). The C-statistic for the best fit
in this case is 4148 with 4064 degrees of freedom. Our upper limit is
roughly consistent with those of \cite{Bulbul12}, who found $\sim
200\kmps$.

Next, we instead used the continuum-subtracted Fe-L profile (bottom
panel of Fig.~\ref{fig:sbprofile}) to create broadened response
matrices, taking the values between dispersion angles of $-0.4$ and
$0.4$ arcmin, where the signal appears significant. A velocity
broadening of $411 \pm 68\kmps$ ($1\sigma$) is obtained using this
dispersion profile. If we discard the negative values after the
continuum subtraction, we obtain a value of $386\kmps$ instead. Both
of these values are consistent with the confidence contours from the
pure RGS spectral fitting shown in Fig.~\ref{fig:vel_size_conf}. The
C-statistic for the best fit with this response matrix is 4103. When
fitting for the spatial broadening (Section \ref{sect:simplefit}), we
obtained the same fit statistic.

\begin{table*}
  \caption{Results using \emph{Chandra} imaging. The \emph{Chandra}
    observations used to make the images are listed, as are the bands
    used to make the line image (Line) and continuum image (Cont.), the radial range
    where the surface brightness are normalised (Norm.) and the
    maximum offset of the profile along the dispersion direction when convolved with the
    response matrix. The background used for the spectral fitting (Bg)
    is either from the observation (O) or from a template (T).
    The velocities shown are limits or measurements
    with $1\sigma$ or 90 per cent uncertainties.
    $^*$Selected background regions were chosen within these radii.}
  \begin{tabular}{lcccccccc}
           System & \emph{Chandra}&Line&Cont.& Norm.       & Max. offset& Bg & Velocity ($\kmps$)  & Velocity ($\kmps$)\\
                   & datasets      &   &     & (arcmin)    & (arcmin)  &     & $1\sigma$           & 90 per cent \\
 \hline
       A\,1068     & 1652          & B & A,C & $0.5-1.0^*$ & $\pm 0.3$ & O   & $<500$              & $<690$ \\
                   &               & B & A,C & $0.5-1.0^*$ & $\pm 0.3$ & T   & $<368$              & $<566$ \\
       A\,1835     & 6880 6881 7370& B & A,C & $0.8-1.4^*$ & $\pm 0.5$ & O   & $<150$              & $<235$ \\
                   &               & B & A,C & $0.8-1.4^*$ & $\pm 0.5$ & T   & $<134$              & $<211$ \\
       A\,2029     & 891 4977      & B & A,C & $0.5-1.3^*$ & $\pm 0.4$ & O   & $<285$              & $<375$ \\
                   &               & B & A,C & $0.5-1.3^*$ & $\pm 0.4$ & T   & $<372$              & $<446$ \\
       A\,2204     & 499 6104 7940 & B & A,C & $0.4-0.6$   & $\pm 0.25$& O   & $278^{+120}_{-135}$ & $<475$ \\
                   &               & B & A,C & $0.4-0.6$   & $\pm 0.25$& T   & $278^{+113}_{-135}$ & $<465$ \\
       A\,2597     & 922 6934 7329 & B & A,C & $0.8-1.2$   & $\pm 0.5$ & O   & $480 \pm 120$       & $480_{-210}^{+190}$\\
                   &               & B & A,C & $0.8-1.2$   & $\pm 0.5$ & T   & $445 \pm 115$       & $445_{-200}^{+175}$\\
                   &               & B & D   & $1.2-1.5$   & $\pm 1.0$ & T   & $330_{-161}^{+128}$ & $<537$ \\
                   &               & B & D   & $2.0-2.5$   & $\pm 1.0$ & T   & $<410$              & $<490$ \\
       A\,3112     & 2216  2516    & B & A,C & $1.4-1.9$   & $\pm 0.4$ & O   & $411 \pm 68$        & $411 \pm 115$\\
                   &               & B & A,C & $1.4-1.9$   & $\pm 0.4$ & T   & $438 \pm 65$        & $438 \pm 110$\\
                   &               & B & A,C & $1.4-1.9$   & $\pm 1.0$ & T   & $337 \pm 80$        & $337_{-139}^{+120}$\\
                   &               & B & A,C & $2.8-3.4$   & $\pm 1.0$ & T   & $331 \pm 81$        & $331_{-141}^{+122}$\\
                   &               & B & A,C & $2.8-3.4$   & $\pm 1.6$ & T   & $350 \pm 80$        & $350_{-138}^{+122}$\\
                   &               & B & D   & $1.4-1.9$   & $\pm 1.0$ & T   & $371 \pm 74$        & $371 \pm 126$\\
                   &               & B & D   & $2.8-3.4$   & $\pm 1.0$ & T   & $277_{-95}^{+81}$   & $277_{-180}^{+130}$\\
                   &               & A,B,C &D& $1.4-1.9$   & $\pm 1.0$ & T   & $373 \pm 74$        & $373_{-126}^{+117}$\\
                   &               & A,B,C &D& $2.8-3.4$   & $\pm 1.0$ & T   & $280_{-95}^{+80}$   & $280_{-180}^{+129}$\\
       E 1455+2232 & 543 4192 7709 & A,B,C&D & $0.5-0.9$   & $\pm 0.3$ & O   & $<180$              & $<304$ \\
                   &               & A,B,C&D & $0.5-0.9$   & $\pm 0.3$ & T   & $<200$              & $<345$ \\
       Hydra A     & 4969 4970     & B & A,C & $0.8-1.6^*$ & $\pm 1.0$ & O   & $<510$              & $<640$ \\
                   &               & B & A,C & $0.8-1.6^*$ & $\pm 1.0$ & T   & $504_{-211}^{+192}$ & $504_{-405}^{+320}$ \\
    MS 2137.3-2353 & 928 4974 5250 & A,B,C&D & $0.3-0.6$   & $\pm 0.2$ & O   & $450 \pm 200$       & $450_{-410}^{+335}$\\
                   &               & A,B,C&D & $0.3-0.6$   & $\pm 0.2$ & T   & $375 \pm 230$       & $<723$ \\
       NGC 1399    & 319  9530     & A,B & C & $0.4-0.6$   & $\pm 0.4$ & O   & $325^{+65}_{-75}$   & $325^{+114}_{-128}$ \\
                   &               & A,B & C & $1.2-1.5$   & $\pm 1.0$ & T   & $358\pm 76$         & $358_{-57}^{+101}$\\
                   &               & A,B & C & $1.6-2.3$   & $\pm 1.0$ & T   & $370\pm 75$         & $370 \pm 125$ \\
                   &               & A   & C & $1.6-2.3$   & $\pm 1.0$ & T   & $356\pm 75$         & $356 \pm 130$ \\
                   &               & A,B & C & $1.6-2.3$   & $\pm 1.5$ & T   & $382\pm 75$         & $382 \pm 120$ \\
       NGC 4636    & 3926  4415    & A,B & C & $1.3-1.5$   & $\pm 1.0$ & O   & $432_{-80}^{+71}$   & $432_{-135}^{+117}$\\
                   &               & A,B & C & $1.3-1.5$   & $\pm 1.0$ & T   & $581_{-60}^{+65}$   & $581 \pm 103$\\
       Zw 3146     & 909  9371     & B & A,C & $0.7-1.7^*$ & $\pm 0.5$ & O   & $<290$              & $<370$ \\
                   &               & B & A,C & $0.7-1.7^*$ & $\pm 0.5$ & T   & $<289$              & $<360$ \\
    \hline
  \end{tabular}
  \label{tab:spatial}
\end{table*}

\subsubsection{Systematic errors}
There are a number of possible systematic errors to consider. Firstly,
the line width we measure is a function of several different emission
lines, each of which may have different spatial broadening (and
intrinsic width). \emph{Chandra} X-ray images cannot separate these
out spectrally. Even when not accounting for spatial extent, the
signal we measure will be some sort of average signal. Secondly, the
radius at which we normalise the continuum may affect the result
(particularly if there still line emission near the normalisation
radius). The band used to calculate the continuum may not be clean
enough, for example if it contains emission lines. Another potential
problem is that the profile we input into the response generator needs
to be truncated at a particular maximum offset, preferably where the
tail of the surface of the brightness is zero but not including too
much noise. Finally, the choice of background spectrum may have some
effect, as in the previous section.

\begin{figure}
  \includegraphics[width=\columnwidth]{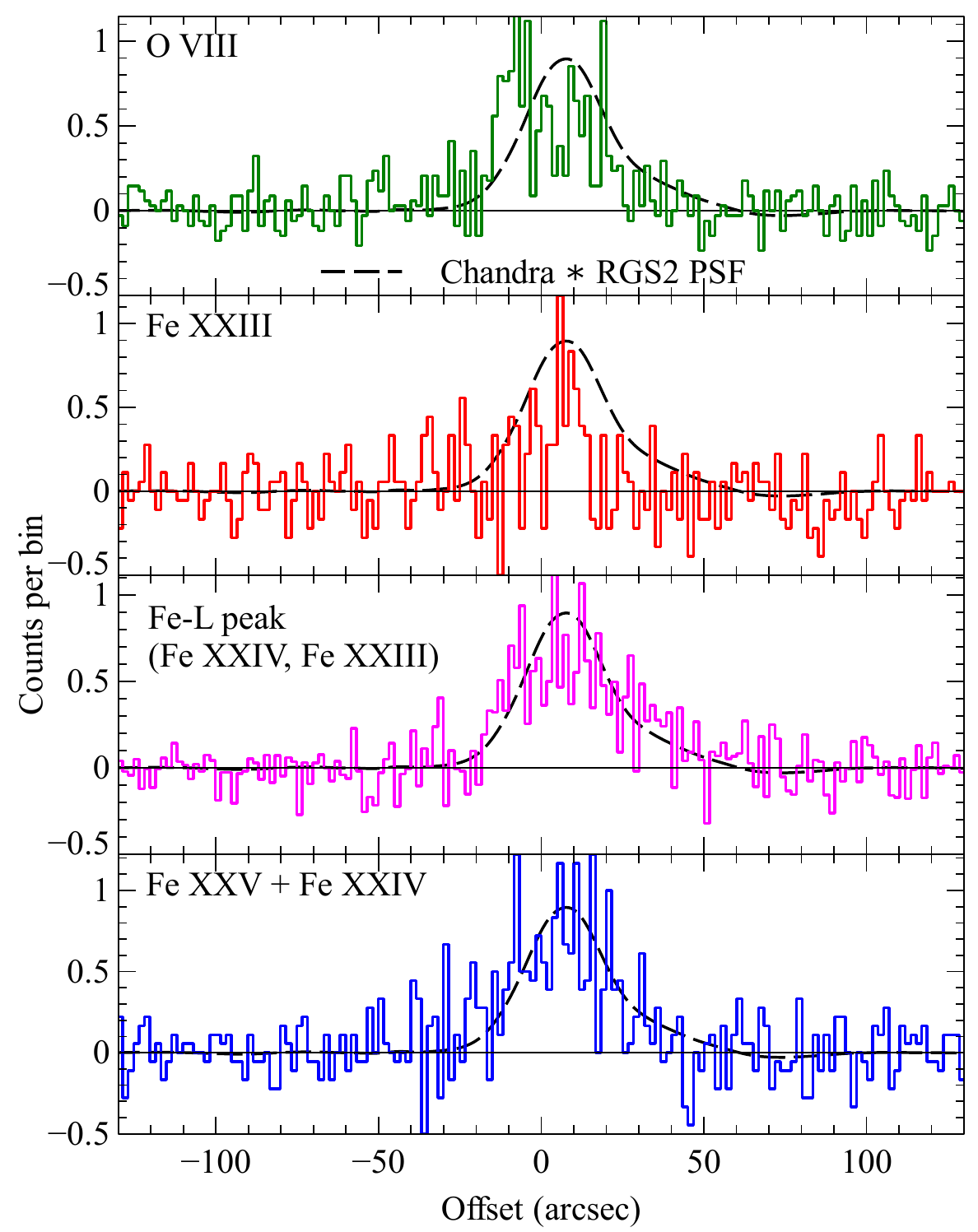}
  \caption{Continuum-subtracted profiles in the cross-dispersion
    directory for various lines from A\,3112. We also plot the
    \emph{Chandra} Fe-L continuum-subtracted profile convolved with
    the RGS2 cross-dispersion PSF at $\beta=0.055$. For
    O~\textsc{viii} in the 1st panel we extract the line from the
    wavelength range 20.35--{20.52\AA} and use both (20.05--20.22,
    20.52--20.69){\AA} as background. For the 2nd panel we use
    12.54--12.71{\AA} as foreground and 12.72--12.90{\AA} as
    background. For the 3rd panel we use 11.75--12.36 and
    (11.37--11.50, 11.62--11.75){\AA} as foreground and background,
    respectively. For the bottom 4th panel, we use 11.32--11.49{\AA}
    as the foreground and 11.14 to 11.31{\AA} as the background.}
  \label{fig:xdisp_profs}
\end{figure}

The \emph{Chandra}-derived continuum-subtracted dispersion axis
profile can be compared to those extracted in the cross-dispersion
direction from the RGS observations themselves. However, unless the
cluster is symmetric they should not necessarily be the
same. Fig.~\ref{fig:xdisp_profs} shows the cross-dispersion surface
brightness profiles in four different lines, subtracting the continuum
from adjacent spectral regions. Also shown on each plot is the
\emph{Chandra} continuum-subtracted Fe-L dispersion axis profile
convolved with the RGS2 cross-dispersion PSF. The plot shows that for
the three strongest lines, there is reasonable agreement with the
\emph{Chandra} profile. For Fe~\textsc{xxiii}, the line appears
narrower than the \emph{Chandra} profile. This is probably because it
is emitted from the cooler material in the cluster, located in the
very central regions. This line is relatively weak and is unlikely to
contribute significantly to the line width signal when fitting the
spectrum.

\begin{figure}
  \includegraphics[width=\columnwidth]{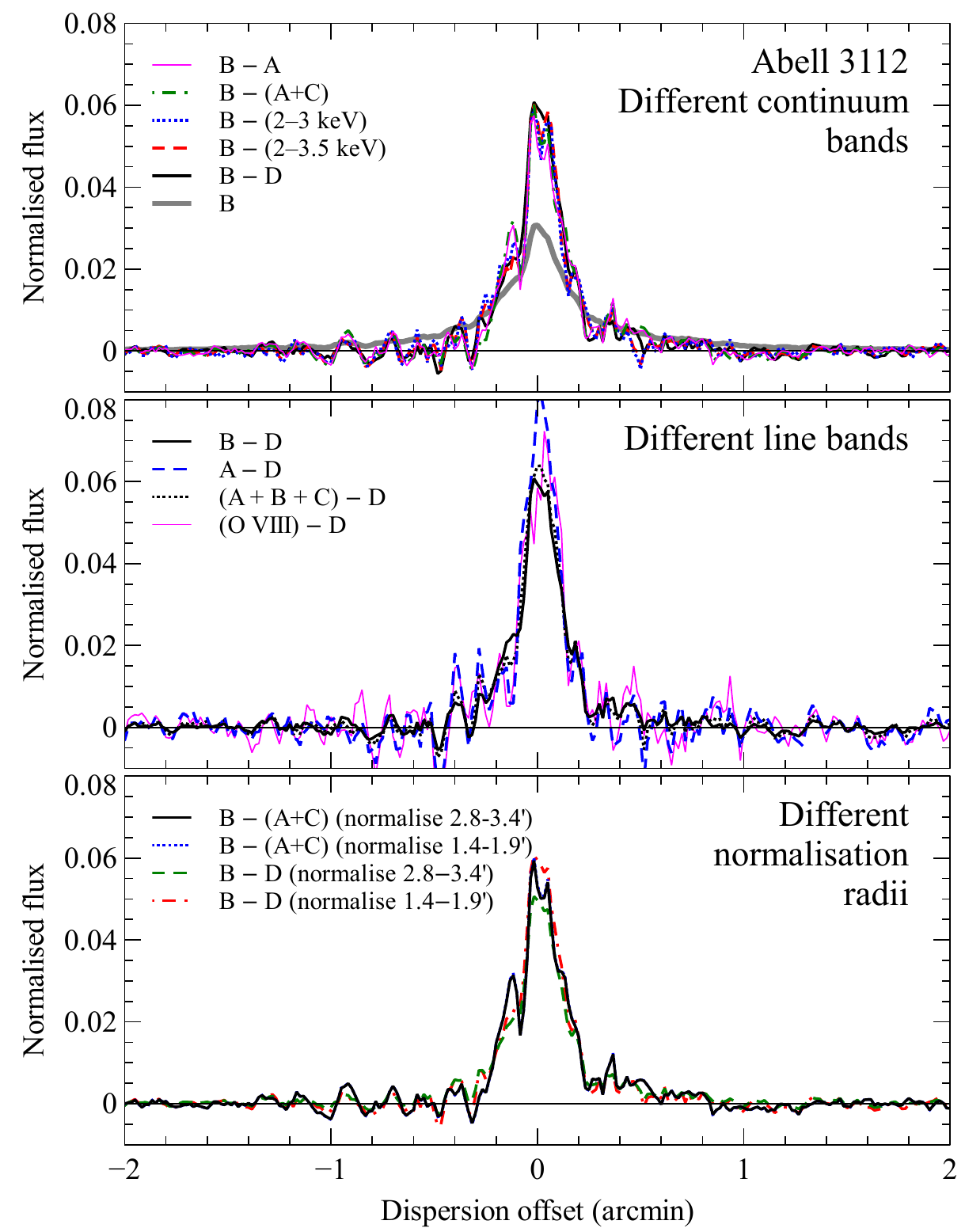}
  \caption{Comparison of continuum subtracted
    profiles. The top panel show the effect of using different
    continuum bands (or none). The centre panel shows the effect of
    using different foreground bands. The bottom panel shows the
    effect of normalising the foreground and background in two
    different radial ranges. The profiles have been normalised to have
    to have a total signal of unity. They have been smoothed using a
    running average, taking the mean of each point with its two
    neighbours. The key shows the line and continuum bands as a
    subtraction.}
  \label{fig:chandraprofiles}
\end{figure}

We can check how the continuum-subtracted profiles from \emph{Chandra}
depend on the choice of spectral bands. Fig.~\ref{fig:chandraprofiles}
shows profiles where different continuum and line bands were used, and
different radii for normalisation of the continuum and line bands. It
can be seen that the continuum band makes little difference to the
continuum-subtracted line profile. The line band has some effect,
where the two softer line bands have stronger peaks in the central
region than the B band profile. The normalisation radius has little
effect when the bands are close in energy (B and A+C), but does affect
the result if they are further away in energy (B and D).

We examined the effect of these possible systematic problems by
repeating the A3112 analysis but varying some of the choices we
made. Using a template background instead of the observation-derived
background increased the best-fitting broadening by $30 \kmps$ in this
object. The line width decreased by $100\kmps$ when the dispersion
axis angular range was increased from $-0.4$ to $0.4$ to $-1$ to $1$
arcmin, using a template background. Increasing the angular range
still further to $1.6$ arcmin increased the line width slightly by $20
\kmps$. If we instead increased the radii where the line and continuum
profiles were normalised from $1.4-1.9$ to $2.8-3.4$ arcmin, this
decreased the velocity by only $6\kmps$. The difference is more
dramatic if a continuum band is chosen which is substantially
different in energy from the line emitting band. Using band $D$ as a
continuum band, the broadening decreased from $370$ to $280\kmps$ when
going from the inner to outer normalisation radii. The inner region
value is close to that when using the adjacent A and C energy bands as
continuum. The effect of the choice of band containing the line
emission can be examined. If we combine the A, B and C bands, using D
as the continuum, there is negligible change to the line width than
when just using B.

These tests show that the broadening we obtain can change by around
$160\kmps$ depending on the choice of bands, background and
radii. This is the likely systematic uncertainty on any results of
broadening. The range of values we obtain for A\,3112 ($180-500\kmps$)
is consistent with the confidence contours shown in
Fig.~\ref{fig:vel_size_conf}.

\subsubsection{Other targets}
We used this technique to include the surface brightness profile for
several targets (listed in Table \ref{tab:spatial}). Unfortunately it
cannot be straightforwardly applied to every object as certain
conditions must be able to be met. Firstly, the core of the centre of
the object must have strong line emission and the outer parts must
not. This is achieved in cool core clusters. It should be unambiguous
where the cool line-emitting region starts and the continuum region
ends. We used \emph{Chandra} temperature maps to help identify where
to normalise the continuum and line profiles. Secondly, if the object
has line emission in different emission lines which are spatially
resolved (for example, due to a temperature gradient over a large cool
core) then the subtraction process often does not work, as the
emission lines move through different bands. This problem is seen in
the Centaurus, A\,2052 and A\,3581 clusters, where the subtracted
profiles are zero at large radii, become negative and then become
positive. Finally, the object should have good \emph{Chandra}
data. The small PSF of \emph{Chandra} means that we do not have to
deconvolve it from the dispersion direction profile and enables us to
easily remove point sources. A future analysis using \emph{XMM} data
should be possible if its PSF is accounted for.

For some distant objects without deep \emph{Chandra} observations the
bands A, B and C are rather narrow and contain too few counts. For
these objects we therefore use A, B and C as foreground bands and a
high-energy D band as a continuum band. For cooler objects, such as
elliptical galaxies, the cool core is at lower temperatures. The Fe-L
peak moves into the A band, instead of the B band. We therefore use
the A and B bands as the foreground bands in these objects and band C
as the background continuum band. The bands used for each observation
are listed in Table \ref{tab:spatial}. As with A\,3112, when
convolving the continuum-subtracted emission line profile with the
response matrices, we trim the profiles in dispersion angle where we
do not see any signal above zero. These angular ranges are also listed
in Table \ref{tab:spatial}.

Our velocity results are listed in Table \ref{tab:spatial}, showing
the statistical uncertainties at the 1$\sigma$ and 90 per cent
confidence levels. These measurements or limits were made using the
$\delta$C approach. The same free parameters were used as when fitting
in the previous section (except for the spatial broadening).

There are a mixture of improved upper limits (such as A\,1835 and
Zw\,3146) and possible detections of broadening of the emission lines
at $300$ to $500 \kmps$ (e.g. A\,3112 and NGC\,4636).

We also investigate some of the possible systematic effects for the
sample. For each object we compare the results using template and
observation-derived spectral backgrounds. In some cases this can make
large changes to the results (e.g. Hydra A, NGC\,4636 and A\,1068),
but for the remainder the effect is small. For NGC\,1399 we
investigate the choice of emission line band, normalisation radius and
extent of the profile included in the convolution. For this object the
range in velocities is around $75 \kmps$, smaller than for A\,3112. We
also examine the choice of continuum in A\,2597 and the maximum
dispersion angle included in the PSF convolution. Similarly to
A\,3112, there is a range of $\sim 150\kmps$ in the velocity
broadening, depending on the choice of parameters.

\subsection{Other methods}
The RGS instruments provide high spectral resolution images of the
source in the cross-dispersion direction. The shape of the source in
the cross-dispersion direction in a particular line can be directly
measured from these images. If the source is close to symmetric in the
dispersion and cross-dispersion directions, then the cross-dispersion
profile can be used to account for the contribution of the extent to
the line width. This method requires that there are enough photons in
the lines for the extent to be measured (See
e.g. Fig.~\ref{fig:xdisp_profs}). For many targets, this will not be
the case for individual lines.

A more complete approach would be to use Monte Carlo modelling of the
cluster to fit all the observations, including imaging and the
cross-dispersion information. \cite{Peterson07} have used smoothed
particles to model galaxy clusters, but have not yet used this
technique to extract velocity width information.

\section{Discussion}
The measurements and limits on the velocity width we present are of
the X-ray coolest gas, as measured using Fe-L and O~\textsc{viii},
characterised by gas at 3 keV or below. When \emph{ASTRO-H}
\citep{Takahashi10} is launched, its microcalorimeter will be most
sensitive to the width of the Fe-K emission lines as it has fixed
spectral resolution in terms of energy. Fe-K emission characterizes
the hotter gas, which may give a wider range of velocities than the
cooler material that the RGS is sensitive to.

It is possible that the coolest gas may have sufficient inertia to
avoid major velocity flows (or it would be destroyed by
them). Previous results using RGS and \emph{Chandra} have shown that
the X-ray coolest gas consists of X-ray cool clumps embedded in a
hotter medium \citep{Sanders2A033509,SandersRGS10}, providing that the
two phases are in pressure balance. \emph{ASTRO-H} will give
complementary data on the hotter, volume-filling component.

The turbulence or velocity structure of the coolest gas may be due to
a number of different physical effects. The active galactic nucleus
(AGN) may be generating kinetic feedback in its environment, including
jets and bubbling \citep{Bruggen05,Heinz10}. Cold fronts, if they
require large flows \citep[e.g.][]{Ascasibar06}, may generate velocity
broadening if they are within the spatial region examined by our
analysis. We note that the physical region probed by our analysis is
different for each of our objects. It is dependent on the redshift of
the object as our extraction region is a fixed angular size and the
physical extent of the line-emitted material within this aperture.

The velocities we measure with RGS are comparable to, or exceed, the
velocity width of optical emission line gas (after converted the
quoted full width at half maximum, FWHM, to $\sigma$). \cite{Hatch07}
examined A\,1068, A\,262, A\,496 and
2A\,0335+096. \cite{CrawfordFabian92} observed the objects AS\,1101,
A\,2597 and A\,496. \cite{Allen92} measured line widths from A\,1835,
Zw\,3146 and A\,1068.

Some of our results are inconsistent with results from previous
analyses of line widths. Our measurement of $280-440 \kmps$ for
A\,3112 does not agree with the upper limit of $200 \kmps$ of
\cite{Bulbul12} when they included the effect of the spatial
distribution of the source. This is probably because the continuum was
not subtracted from the surface brightness profile when computing the
spatial broadening effect. In \cite{Sanders10_Broaden} we used
\emph{Chandra} spatially resolved maps to model expected line widths,
to subtract from observed line widths. This analysis appears to have
overestimated the spatial broadening effect in several objects. This
can be seen by the disproportionate number of objects where the
predicted line width is wider than he observed line width, although
some would be expected if the intrinsic velocity broadening is low.
One likely contributing factor is the finite bin size of the input
maps used to create the simulated spectra. In our previous paper the
spatial and implied velocity broadening were also not added in
quadrature to obtain the observed broadening. Our new method avoids
the problems caused by our previous modelling using spectral maps.

The velocities we obtain for NGC\,5044 and NGC\,5813 are consistent
with those found by \cite{dePlaa12} using both line widths and
resonance scattering.

We note that for NGC\,4636, the $430-580 \kmps$ broadening we measure
is apparently inconsistent with the $100\kmps$ upper limit inferred
from the presence of resonance scattering \citep{Werner09}. However,
these values may still be in agreement if the broadening we measure is
the result of ordered motion. The radio source is likely to produce a
bipolar outflow which will give a strong resonance scattering signal
due to the low relative velocities of gas on either side. The emission
lines, however, would be broadened by the difference in outflow
velocities between each side.  Alternatively, the disagreement may be
because the spatial distribution of the line emission is not properly
accounted for in our analysis. Although the \emph{Chandra} surface
brightness profile appears consistent with the location of the
brightest and coolest part of the group ($\sim 0.6\keV$ temperature,
in a 0.3 arcmin radius core, with 1 arcmin wings), the profiles of
individual lines may be different in this rather complex
source. Another possible uncertainty is that the resonance scatter
results may be affected theoretical uncertainties on the relative
strengths of the Fe~\textsc{xvii} emission lines as a function of
temperature \citep[see e.g.][]{SandersRGS08}.

In Section \ref{sect:simplefit} we measured the velocities assuming
that the line surface brightness could be fitted well by a Gaussian
model. There will be a systematic uncertainty associated with this
assumption. In the case of A\,3112 this assumption is a reasonable
one, as shown by the continuum-subtracted surface brightness profile
in Fig.~\ref{fig:sbprofile} and cross-dispersion line profiles in
Fig.~\ref{fig:xdisp_profs}. The line width results for A\,3112 from
this method (Table \ref{tab:mainresults} and
Fig.~\ref{fig:vel_size_conf}) also agree well with the results using
the technique using the \emph{Chandra} surface brightness profiles
(Table \ref{tab:spatial}). Other objects with resolved or
marginally-resolved line widths show good agreement between the
methods, including NGC 1399, MS\,2137, A\,2204 and A\,2597. The object
with the largest difference between the methods is NGC\,4636. This
target has a complex morphology so the Gaussian assumption may not be
good in this case.

Detailed spatially-resolved line profiles (measuring line widths and
positions as a function of radius) will enable \citep{Zhuravleva12}
the turbulent spectrum of the velocity field to be determined. We
await the results from future X-ray missions flying high spectral
resolution detectors, such as \emph{ASTRO-H}, to examine line shapes
in detail as a function of position.

In our analysis, we used both a conventional and a MCMC method for
examining uncertainties on parameters. For most of our objects, the
results agreed. The use the affine-invariant MCMC sampler made it much
easier to do the MCMC analysis compared to Metropolis-Hastings. With
Metropolis-Hastings it is often difficult to choose an appropriate
proposal distribution, leading to the parameter space being poorly
sampled because of too many or too few rejections. The algorithm also
has difficulty when parameters are strongly correlated. The
affine-invariant MCMC analysis was not a complete success, however,
because in several cases we had to increase the number of walkers to
get agreement between the conventional and MCMC results. Some parts of
parameter space were missed in the MCMC analysis for some
objects. Therefore, it was much easier to get quick, robust and stable
results using conventional techniques rather than MCMC.  The advantage
of the MCMC approach was that we were able to examine the correlations
of each parameter with every other and with the fit statistic
(e.g. Fig.~\ref{fig:vel_size_conf}). This allowed us to understand the
modelling better. In a few cases, there are still divergent results
between MCMC and the conventional technique, which may indicate that
the C-statistic uncertainties are underestimated because of a tail in
the posterior probability distribution.

\section{Conclusions}
We present limits or measurements of the velocity width of the soft
X-ray emission for a sample of clusters and groups of galaxies, and
elliptical galaxies, using data from \emph{XMM} RGS. In our analysis
we take two different approaches to including the spatial broadening
intrinsic to the RGS instruments. For all targets, we model the
spatial broadening with a Gaussian component in the spectral
modelling. For a subset of objects we also use a second approach where
\emph{Chandra} imaging data in a continuum-subtracted line-emitting
band is used to model the spatial broadening.

Using Gaussian spatially modelling, for twenty of the targets we find
upper limits on the velocity broadening of less than $500\kmps$ and
find five targets with limits of $300\kmps$ or lower. There are some
targets for which we detect velocity broadening with 90 per cent
confidence between $300$ to $500\kmps$, including A\,3112, HCG\,62 and
NGC\,1399. NGC\,4636 shows broadening if \emph{Chandra} imaging is
used in the spectral analysis.

\section*{Acknowledgements}
ACF acknowledges the support of the Royal Society. We thank Randall
Smith for supplying the APEC code. We are grateful to an anonymous
referee for comments which significantly improved this paper.

\bibliographystyle{mnras}
\small
\bibliography{refs}

\end{document}